\documentclass{ws-procs975x65}
\usepackage{graphicx}

\begin{document}

\title{Pulsar observations with European telescopes for testing gravity and detecting gravitational waves}
\author{Delphine Perrodin$^{1*}$, Cees G. Bassa$^2$, Gemma H. Janssen$^2$, Ramesh Karuppusamy$^3$, Michael Kramer$^3$, Kejia Lee$^4$, Kuo Liu$^3$, James McKee$^5$, Mark Purver$^5$, Sotiris Sanidas$^6$, Roy Smits$^3$, Benjamin W. Stappers$^5$, Weiwei Zhu$^3$, Raimondo Concu$^1$, Andrea Melis$^1$, Marta Burgay$^1$, Silvia Casu$^1$, Alessandro Corongiu$^1$, Elise Egron$^1$, Noemi N. Iacolina$^1$, Alberto P. Pellizzoni$^1$, Maura Pilia$^1$, Alessio Trois$^1$}

\address{$^1$INAF - Osservatorio Astronomico di Cagliari,\\
Via della Scienza 5, 09047 Selargius (CA), Italy\\
$^*$E-mail: delphine@oa-cagliari.inaf.it}
\address{$^2$ASTRON, the Netherlands Institute for Radio Astronomy,\\
Postbus 2, 7990 AA, Dwingeloo, The Netherlands}
\address{$^3$Max Planck Institut f{\"u}r Radioastronomie,\\
Auf dem H{\"u}gel 69, 53121 Bonn, Germany}
\address{$^4$Kavli Institute for Astronomy and Astrophysics, 
Peking University, \\ Beijing 100871, P. R. China}
\address{$^5$Jodrell Bank Centre for Astrophysics, The University of Manchester,
\\ Manchester, M13 9PL, United Kingdom}
\address{$^6$Anton Pannekoek Institute for Astronomy, University of Amsterdam,
\\ Science Park 904, 1098 XH Amsterdam, The Netherlands}

\begin{abstract}
A background of nanohertz gravitational waves from supermassive black hole binaries could soon be detected by pulsar timing arrays, which measure the times-of-arrival of radio pulses from millisecond pulsars with very high precision. The European Pulsar Timing Array uses five large European radio telescopes to monitor high-precision millisecond pulsars, imposing in this way strong constraints on a gravitational wave background. To achieve the necessary precision needed to detect gravitational waves, the Large European Array for Pulsars (LEAP) performs simultaneous observations of pulsars with all five telescopes, which allows us to coherently add the radio pulses, maximize the signal-to-noise of pulsar signals and increase the precision of times-of-arrival. We report on the progress made and results obtained by the LEAP collaboration, and in particular on the addition of the Sardinia Radio Telescope to the LEAP observations during its scientific validation phase. In addition, we discuss how LEAP can be used to monitor strong-gravity systems such as double neutron star systems and impose strong constraints on post-keplerian parameters.
\end{abstract}


\bodymatter

\section{Pulsar Timing Arrays for Gravitational Wave Detection}

Exactly a hundred years after Einstein's formulation of general relativity, we are on the verge of finding the ``holy grail" that would open a new window on the universe: gravitational waves (GWs). In addition to Earth-based interferometers such as LIGO and VIRGO, and the future space-based eLISA, pulsar timing arrays (PTAs) are actively searching for GWs using millisecond pulsars (MSPs) and Earth as test masses. GWs affect the space-time between Earth and pulsars, and introduce offsets in the times-of-arrival (TOAs) of radio pulses emitted by pulsars. Due to their long baselines (Earth-pulsar distances), PTAs are sensitive to GWs in the nanohertz frequency range, making it a complementary tool to LIGO and the planned eLISA. This frequency range includes the predicted GW emission from supermassive black hole binaries and cosmic strings. For the purpose of GW detection, PTAs monitor the timing residuals of MSPs (defined by the the difference between expected and observed TOAs) over a long period ranging from 10 to 30 years. A detection can be achieved by studying the timing residuals of a number of different pulsars and finding specific correlations in the timing residuals between pulsar pairs.
In order to estimate any ``disturbance" in the arrival times of radio pulses, we need to account for every rotation of the pulsar. In a typical PTA observation, we observe a pulsar for minutes to hours; correct for interstellar dispersion and fold on the known pulsar period;  determine TOAs by correlating the obtained profile with a template profile; improve the pulsar timing model; study timing residuals. GW detection pipelines can then be applied to achieve a detection or set an upper limit to a background of GWs.\cite{EPTA}

Current PTAs include the North American Nanohertz Observatory for Gravitational Waves (NANOGrav), the European Pulsar Timing Array (EPTA) and the Parkes Pulsar Timing Array (PPTA), which collaborate to form the International Pulsar Timing Array (IPTA). Here we will focus on the EPTA's efforts to detect GWs using high-precision pulsar timing at five large radio telescopes: the 100-meter Effelsberg telescope in Germany; the 94m-equivalent Nan{\c c}ay Radio Telescope in France; the 94m-equivalent Westerbork Synthesis Radio Telescope (WSRT) in the Netherlands; the 76-m Lovell Telescope in the UK, and the 64-m newly-commissioned Sardinia Radio Telescope (SRT) in Italy. 

\section{LEAP Project Overview}

As part of the EPTA's efforts to detect GWs, the Large European Array for Pulsars (LEAP) project conducts simultaneous observations of MSPs with the five 100-m class European telescopes, which allows us to add the pulsar data from each telescope coherently and improve TOA precision. By conducting simultaneous observations, the telescopes are combined into a tied array, effectively creating a single, fully-steerable telescope with the equivalent size of a 195-m dish, which is comparable to the aperture of the illuminated Arecibo dish with a large range of declinations (-30 to 90 degrees). The large aperture improves TOA precision and enables the timing of weaker pulsars. In a tied-array telescope, the signals from different telescopes are corrected for differences in time delay, then added in phase. These time delays can be due to differences in geometry, observatory clocks, instruments or atmospheric conditions. Therefore in addition to improving TOA precision, the LEAP project also helps to calibrate the instrument delays between telescopes and spot any anomalous offsets (``jumps") at one of the telescopes.

The LEAP project officially started in 2009, and routine, monthly observations started in early 2012 with three telescopes (Effelsberg, Lovell and WSRT). The Nan{\c c}ay telescope was added in the middle of 2012, while SRT joined with one 16-MHz sub-band in July 2013 and with the full bandwidth in March 2014 during the telescope's scientific validation phase. Each monthly session is 25 hours in length and typically monitors 22-23 MSPs simultaneously at all five telescopes; both pulsars and phase calibrators are observed. The observing is done at L-band,  with a center frequency of 1396 MHz and a bandwidth of 128 MHz. Baseband data (corresponding to raw voltages) are recorded at each telescope and saved to disk. Disks are later shipped to Jodrell Bank Observatory for correlations. The LEAP pipeline designed to correlate the data was developed from scratch for the purpose of coherent addition. Calibrator and pulsar data are correlated to obtain time and phase offsets between telescopes (with the process of ``fringe finding"). Once these offsets are found, they are applied to the pulsar data, which are then added coherently to maximize S/N. Added LEAP baseband files are obtained, from which we generate archives and TOAs using standard pulsar software. This pipeline has many applications. It can for example be used to add other telescopes, such as in the case of the J1713+0747 global campaign (GiantLEAP). The LEAP project is effectively a precursor to SKA science and provides a TOA precision comparable to SKA Phase 1. A detailed overview of the LEAP project is found in Ref.~\refcite{LEAP}.

\section{LEAP Project Implementation at the Sardinia Radio Telescope}

The Sardinia Radio Telescope was completed in 2011 and is a fully-steerable 64-meter dish with three main focal positions, a wide planned frequency range (300 MHz to 100 GHz) and an active surface. It has been undergoing its scientific validation phase.\cite{AV}. Its active participation in the VLBI and LEAP international collaborations is described in Ref.~\refcite{EVN}. SRT is the latest addition to the LEAP project. The addition of one telescope leads to a huge increase in sensitivity since the LEAP data are coherently added: in fact, the S/N increases linearly with the number of telescopes (for identical telescopes), as opposed to the square root of the number of telescopes in the case of incoherent addition. 

The dual L/P receiver was installed at the primary focus of the telescope in the spring of 2013. A ROACH backend \footnote{https://casper.berkeley.edu/wiki/ROACH} was installed at the site in July 2013 and, using the PSRDADA software \footnote{http://psrdada.sourceforge.net}, was set up to record baseband data to disk. SRT joined LEAP for the first time in July 2013 for a single 16-MHz sub-band (1428-1444 MHz), and observations were repeated monthly.  In February 2014, an 8-node CPU cluster was installed to allow the simultaneous recording of baseband data in 8 16-MHz channels (16 MHz per node), therefore covering the full LEAP band. Starting in March 2014, SRT joined monthly LEAP runs with the full bandwidth for 20 pulsars. Data acquisition during LEAP runs was subsequently fully automated.
The LEAP storage cluster, with 96 TB of storage, was installed in April 2014. It is able to store the 40 TB of LEAP data taken at SRT each month. After each run, data are copied from the CPU cluster to the storage cluster and disks are shipped to Jodrell Bank Observatory. Standard pulsar software (PSRCHIVE and TEMPO2) were upgraded to enable data analysis with SRT data (PSRCHIVE was upgraded as of August 2014 \footnote{thanks to Willem van Straten} and TEMPO2 as of May 2015 \footnote{thanks to George Hobbs}).

The first fringes using the LEAP correlating pipeline were found during a test between SRT and WSRT in May 2014, as seen in Fig. 1. The 5-telescope data are now correlated on a monthly basis at Jodrell Bank. The SRT data however lags in quality compared to the other telescopes because of the presence of strong Radio Frequency Interference. In particular, a malfunctioning military radar is severely impacting the quality of the entire LEAP bandwidth.

\vspace{0.4cm}
\begin{figure}
\begin{center}
\includegraphics[width=2.5in]{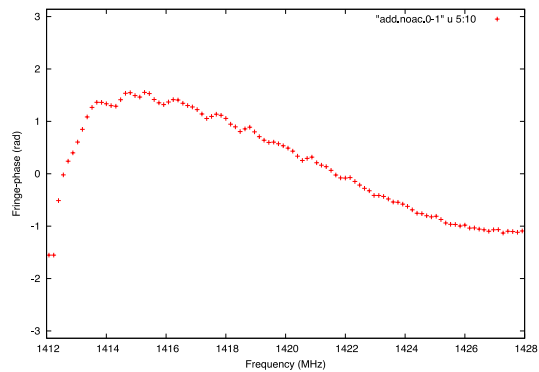}
\end{center}
\caption{First fringe found between SRT and WSRT using the LEAP correlation pipeline: 5 seconds of quasar 3C454 at 1420 MHz (May 2014)}
\end{figure}

\section {LEAP First Results}
The data acquisition hardware and software are now in place at all telescopes, including SRT, and all five telescopes are fully participating in monthly LEAP sessions. The LEAP reduction pipeline is now up-to-date and the data reduction is done on a month-to-month basis, while a backlog of existing data is being reduced in parallel. 
For 100$\%$ coherence, we expect the S/N of the added LEAP data to be equal to the sum of the S/N of the individual telescopes. Coherence is generally achieved in the strongest pulsars such as J1022+1001, as shown in Fig. 2. 

\begin{figure}
\begin{center}
\includegraphics[width=1.5in,angle=-90]{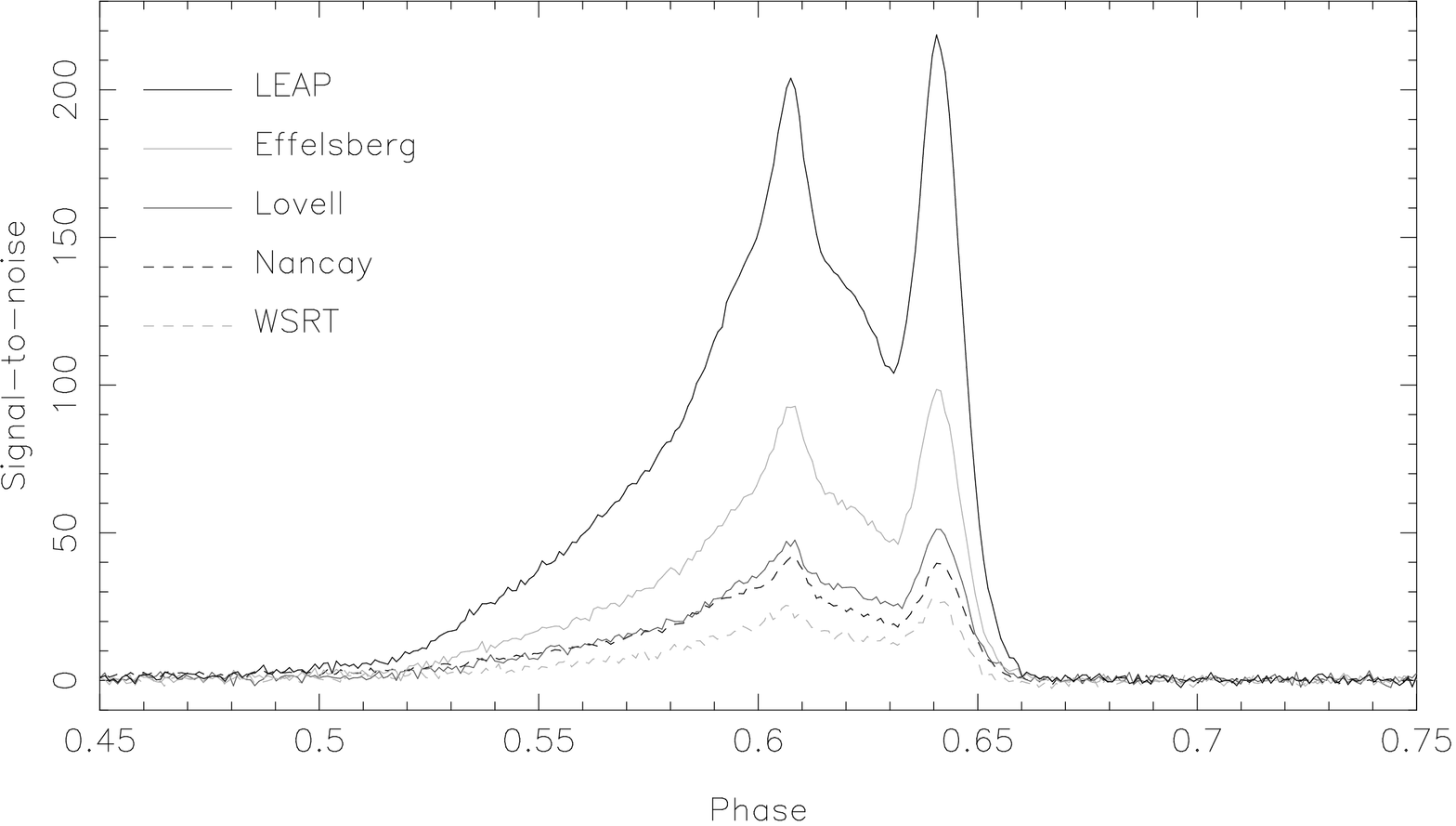}
\includegraphics[width=1.5in,angle=-90]{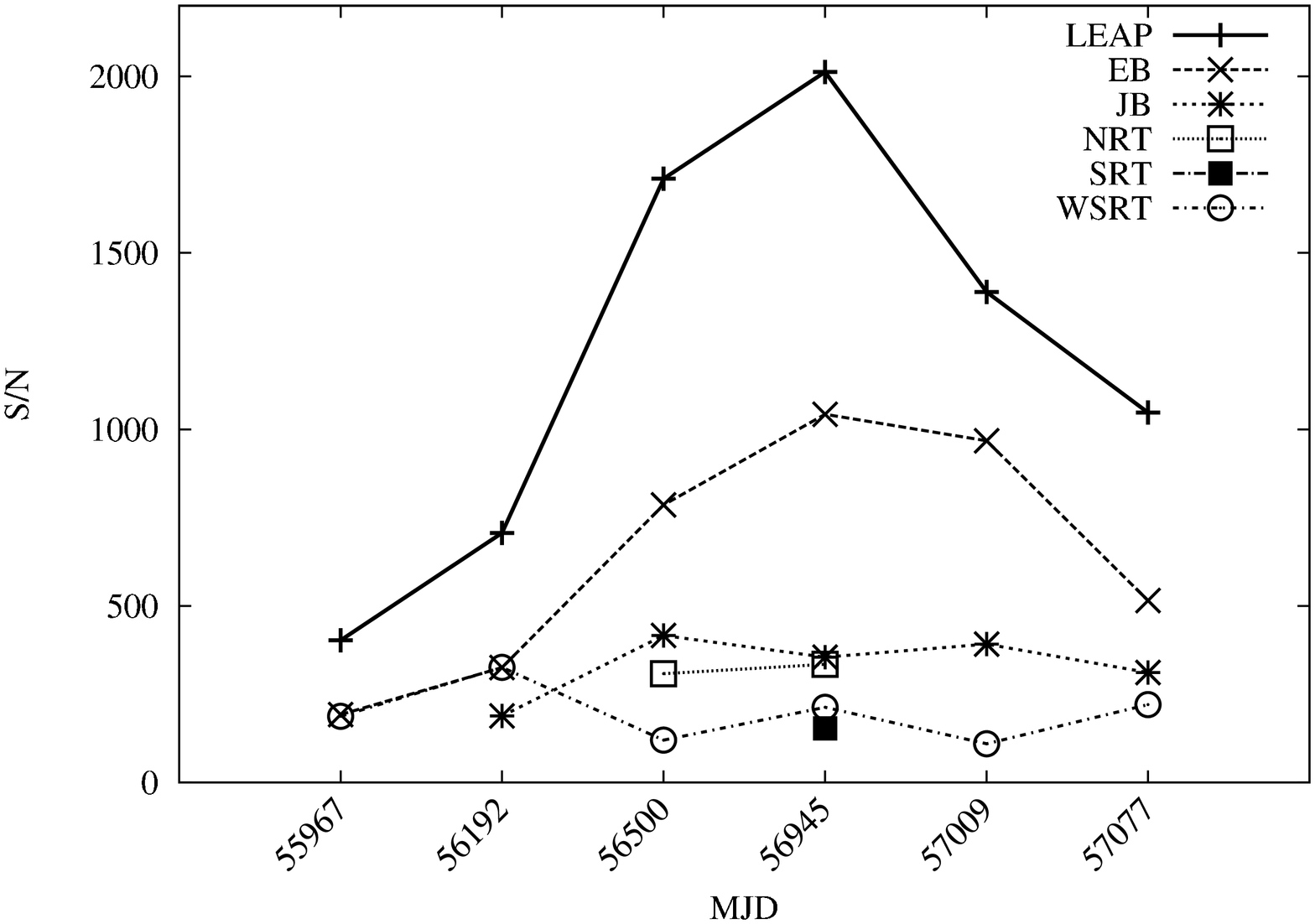}
\vspace{0.2cm}
\caption{Coherence of added LEAP data. On the left, pulsar profiles of PSR J1022+1001 at MJD 56500 for individual telescope data and for the 4-telescope addition (Effelsberg, Lovell, Nan{\c c}ay, WSRT). On the right, S/N for individual telescopes and coherently-added data for different epochs from February 2012 until February 2015. The pulsar shows near perfect coherence, in that the LEAP S/N is roughly equal to the sum of the S/N of the individual telescopes.}
\end{center}
\end{figure}

A higher S/N in the pulsar profiles leads to an improvement in timing precision, i.e. we are better able to constrain pulsar models and ultimately, find gravitational waves. We show an example of increased timing precision in Fig. 3.
\begin{figure}
\begin{center}
\includegraphics[width=2in,angle=-90]{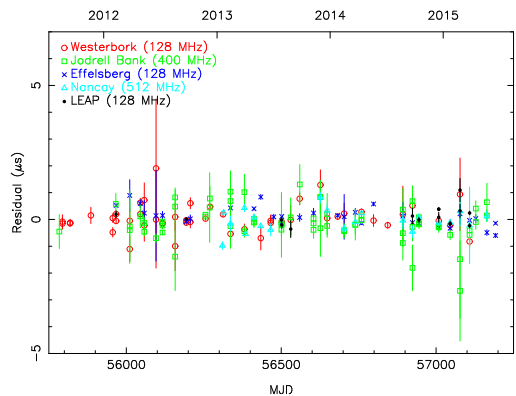}
\end{center}
\caption{Timing residuals of PSR J1713+0747 for single telescope data as well as coherently-added LEAP data over a five-year timespan. The data from the individual telescopes (including the LEAP data) have a residual rms of 0.25 $\mu s$, while the LEAP data alone has an rms of 0.18 $\mu s$.}
\end{figure}

By improving the S/N of pulsar profiles, LEAP improves pulsar timing precision (i.e. lowers the rms of timing residuals), which is necessary for extracting a gravitational wave signal. However, LEAP can also be used to study phase jitter (Liu et al, in preparation) or perform pulsar searching for weak pulsars with known positions. Details are found in Ref.~\refcite{LEAP}.

\section{Tests of Strong Gravity in Double Neutron Star Systems}
Double neutron star (DNS) systems provide great tests of strong gravity. In particular, DNS systems help us constrain post-keplerian parameters, such as the rate of advance of periastron, the Einstein delay, the Shapiro delay, and the orbital period decay. The constraints on these parameters can be significantly improved by observing DNS systems with LEAP. Indeed, the coherent addition of pulsar data with LEAP significantly increases the S/N of pulsar observations, leading to a lower residual rms and better constraints on the pulsar model. We are planning a campaign to monitor DNS sources on a monthly basis during LEAP runs for the purpose of testing strong gravity (Perrodin et al, in preparation).

In the pulsars that are known DNS sources, one can also search for a pulsar companion, discovering new double-pulsar systems, which would provide excellent tests of strong gravity. In fact, one of the many applications for LEAP is pulsar searching. Thanks to the increased sensitivity of the LEAP tied-array beam, LEAP could search for new double-pulsar systems. As a test, we performed a blind search on 5 minutes of coherently-added LEAP data of the known double neutron star PSR J1518+4904 with Effelsberg and WSRT, searching for the pulsation of the neutron star companion. 33 candidates were found with the same DM as J1518+4904, but were found to be harmonics of the pulsar or RFI. J1518+4904 is observed monthly with LEAP, and we can continue to search for a pulsar companion in the added data.

\section{Conclusion}
PTAs aim to detect nanohertz GWs from supermassive black hole binaries. For this purpose, LEAP achieves higher sensitivity, leading to higher constraints on a GW background from supermassive black hole binaries. The recent addition of SRT to LEAP increases the sensitivity of the ``LEAP telescope". LEAP can also be used to constrain the strong gravity environment of pulsars in double neutron star systems.

\section*{Acknowledgments}

The authors acknowledge the support of colleagues in the EPTA collaboration. The presented work has been funded by the ERC Advanced Grant ``LEAP", Grant Agreement Number 227947 (PI M. Kramer). The work at the Sardinia Radio Telescope, which is operated by the Istituto Nazionale di Astrofisica (INAF), was done during the scientific validation phase of the telescope. We thank the SRT Astrophysical Validation Team \footnote{http://www.srt.inaf.it/astronomers/astrophysical-validation-team/}.

\end{document}